\documentclass[preprint,
aps,superscriptaddress,longbibliography,prb]{revtex4-2}

\usepackage{amssymb}
\usepackage{amsmath}
\usepackage{amsfonts}
\usepackage{bm}
\usepackage{graphicx}
\usepackage{color}
\usepackage[usenames,dvipsnames]{xcolor}
\usepackage[displaymath]{lineno}

\usepackage{titlesec}
\titleformat{\section}
  {\large\bfseries\raggedright}{\thesection.}{1em}{}[]
\titleformat{\subsection}
  {\normalfont\bfseries\raggedright}{\thesubsection.}{1em}{}[]
\titlespacing*{\section}{0pt}{*3}{*1}
\titlespacing*{\subsection}{0pt}{*2}{*0.5}

\usepackage{hyperref}
\hypersetup{
	colorlinks,
	linkcolor={blue},
	citecolor={blue},
	urlcolor={blue}
}

\usepackage[utf8]{inputenc}
\usepackage[T1]{fontenc}     
\usepackage{lmodern}
\usepackage{multirow}
\usepackage{epsfig}
\usepackage{url}
\usepackage{xcolor}
\usepackage[normalem]{ulem}
\usepackage{comment}

\usepackage{tabularx}
\usepackage{sidecap}
\usepackage{soul}
\usepackage{float}

\usepackage{xr-hyper} 
\usepackage{import}
\usepackage{orcidlink}

\usepackage{array}
\allowdisplaybreaks

\usepackage[capitalize]{cleveref}
\usepackage{textgreek}

\crefname{figure}{Fig.}{Figs.}

\begin{document}

\title{Emergence of spin entanglement with the pseudogap onset in the Fermi-Hubbard model}

\author{Frederic Bippus \orcidlink{0009-0006-4316-6547}}
\affiliation{Institute of Solid State Physics, TU Wien, 1040 Vienna, Austria}

\author{Thomas Chalopin \orcidlink{0000-0001-7633-0442}}
\affiliation{Université Paris-Saclay, Institut d’Optique Graduate School, CNRS, Laboratoire Charles Fabry, Palaiseau 91127, France}
\affiliation{Max-Planck-Institut für Quantenoptik, Garching 85748, Germany}
\affiliation{Munich Center for Quantum Science and Technology, Munich 80799, Germany}

\author{Gabriele Bellomia~\orcidlink{0000-0003-1576-3388}}
\affiliation{Institute of Solid State Physics, TU Wien, 1040 Vienna, Austria}
\affiliation{Scuola Internazionale Superiore di Studi Avanzati (SISSA), 34136 Trieste, Italy}

\author{Gerg\H o Ro\'osz \orcidlink{0000-0002-9541-119X}} 
\affiliation{Institute of Solid State Physics, TU Wien, 1040 Vienna, Austria}

\author{Titus Franz \orcidlink{0009-0007-8363-2896}}
\affiliation{Max-Planck-Institut für Quantenoptik, Garching 85748, Germany}
\affiliation{Munich Center for Quantum Science and Technology, Munich 80799, Germany}

\author{Antoine Georges \orcidlink{0000-0001-9479-9682}}
\affiliation{Collège de France, Paris Sciences et Lettres University, Paris 75005, France}
\affiliation{Center for Computational Quantum Physics, Flatiron Institute, New York, NY 10010}
\affiliation{École Polytechnique, Centre de Physique Théorique, CNRS, Paris F-91128, Palaiseau, France}
\affiliation{Department of Quantum Matter Physics, Université de Genève, Genève CH-1211, Suisse}

\author{Anna Kauch \orcidlink{0000-0002-7669-0090}}
\affiliation{Institute of Solid State Physics, TU Wien, 1040 Vienna, Austria}

\author{Immanuel Bloch \orcidlink{0000-0002-0679-4759}}
\affiliation{Max-Planck-Institut für Quantenoptik, Garching 85748, Germany}
\affiliation{Fakultät für Physik, Ludwig-Maximilians-Universität, Munich 80799, Germany}
\affiliation{Munich Center for Quantum Science and Technology,
Munich 80799, Germany}

\author{Karsten Held \orcidlink{0000-0001-5984-8549}}
\affiliation{Institute of Solid State Physics, TU Wien, 1040 Vienna, Austria}

\maketitle
\begin{widetext}
{\bf Despite decades of intense theoretical and experimental investigation, the two-dimensional Fermi-Hubbard model still resists a complete microscopic understanding. Conventional approaches typically probe global observables and locally resolved correlation functions. Here, we develop a complementary perspective 
based on the measurement of entanglement. Using both an ultracold-atom quantum simulator and numerical simulations based on the dynamical vertex approximation, we find that entanglement is closely tied to the onset of the enigmatic pseudogap regime: spin-singlet entanglement emerges only as the pseudogap sets in and, in contrast to classical correlations, remains confined to nearest-neighbour sites in this regime. Our results, therefore, disfavour purely classical-fluctuation theories of the pseudogap and constrain microscopic models to those that develop nearest-neighbour spin-singlet entanglement at the pseudogap onset.}
\end{widetext}

Correlations can be classical or quantum, as exemplified {on the one hand} by the classical magnetic correlations of the Ising model and {on the other hand} by spin-singlet quantum correlations. The textbook example of the latter is the two-site spin singlet, and one possible extension to the lattice is Anderson's \cite{Anderson_1973,Anderson1987} resonating valence bond state. The static
magnetic susceptibility alone cannot distinguish between classical and quantum correlations, and indeed both fall in the same finite-temperature universality class whenever the dimension and the order-parameter symmetry of the underlying transition coincide. In contrast, entanglement only captures quantum correlations. It therefore offers a complementary microscopic perspective on correlated Fermi systems.

Hitherto, entanglement studies have been primarily focused on spin systems~\cite{Zoller_2023_XXZ_entanglement,Zoller_2021_entanglement_hamiltonian,Amico_entanglement,Subrahmanyam_2004}, natural realizations of coupled quantum bits (qubits), and bosonic systems \cite{islam_measuring_2015}. Much less is known, both experimentally and theoretically, about entanglement in correlated Fermi systems at finite temperatures: Besides the second Rényi entropy \cite{grover_entanglement_2013}, which has even been simulated on a quantum computer \cite{Monroe_quantum_computer_Entanglement_2018}, and the mutual information \cite{Cocchi_2017,Walsh_MI_2019,Noak_momentum_space_entanglement_2015}, the quantum Fisher information \cite{hauke_measuring_2016,Laurell_2025} has been at the focus of recent research. 
It is obtained from the dynamical magnetic susceptibility, and has been measured for antiferromagnetic spin chains~\cite{Mathew2020,laurell_quantifying_2021}, quantum spin liquids \cite{Scheie2024}, heavy fermion systems \cite{mazza_quantum_2024}, and cuprates \cite{balut2025quantumfisherinformationreveals,Bippus_2025_pseudogap_QFI}.
While readily accessible by neutron scattering, the quantum Fisher information only constrains the entanglement depth – how many constituents are entangled, but not which ones, nor where in the system the entanglement resides.

Spatial information and the nature of entanglement can be studied, on the other hand, from the two-site reduced density matrix at varying distance between the two sites \cite{Roosz_2024_2sRDM,Bippus_2sRDM_entanglement_2026}.
The $16\times16$  components of the reduced density matrix can be further reduced if we restrict our focus to Alice--Bob-type of entanglement between the two sites, where Alice only measures observables on one of the two sites and Bob on the other. Such measurements then conserve charge and parity on Bob's and Alice's respective sites \cite{Ding_2024_SSR,local_qinfo_Hubbard,BellomiaPhD,Bellomia2023}.
The superselection rules (SSR) single out precisely the component of the entanglement that is operationally accessible to quantum-information protocols under such restrictions~\cite{Banhuls2007,Friis_SSR_required_by_relativity_2016,Szalay2021}.

In this paper, we measure all components of the SSR two-site reduced density matrix in a Fermi-Hubbard ultracold atom quantum simulator \cite{Bloch_QSIM_revieew,Chalopin_optical_superalattices_2025} and compare them to our calculations based on the dynamical vertex approximation (D$\Gamma$A) \cite{Toschi2007,rohringer_diagrammatic_2018}. Our main finding is summarized in Fig.~\ref{Fig:main}. Across most of the temperature $T$ and doping $\delta$ phase diagram no entanglement is detected, signalled by a vanishing SSR
negativity (black). Only with the onset of the pseudogap phase (white solid line) at lower temperatures and moderate doping, entanglement sets in. Experiment and D$\Gamma$A are in good agreement across the sampled $T,\delta$ range.

Before discussing our results in detail, let us briefly put the two methods used in perspective: Given the numerical difficulty of accessing the Fermi-Hubbard model in its most physically relevant parameter regimes, ultracold-atom quantum simulators provide a powerful experimental alternative. These quantum simulators have been especially useful in providing novel insight on the microscopic spin and charge correlations \cite{parsons:2016a, boll:2016a, cheuk:2016a, koepsell:2019, hirthe:2023,Zhu2025}, as well as response functions \cite{brown:2019a, prichard:2025} in the Fermi-Hubbard model, including its pseudogap phase \cite{Greiner_PG_2025,
Chalopin_PG_2026}. From the correlation functions of these quantum simulators, we here extract the operationally accessible entanglement. Our theoretical calculations are, on the other hand, based on D$\Gamma$A \cite{Toschi2007,rohringer_diagrammatic_2018}, a diagrammatic extension of dynamical mean-field theory \cite{Georges1996}, which has been found to provide an excellent description of antiferromagnetic fluctuations, the pseudogap \cite{Schaefer2021}, and even predicted \cite{Kitatani2020} the superconducting phase diagram of nickelate superconductors \cite{lee2023linear}.

\section*{SSR entanglement}
Consider two subsystems controlled by Alice (A) and Bob (B). Entanglement is the non-classical part of the information about B that can be acquired through a measurement on A. In this bipartite setting, the (reduced) density matrix $\rho_{AB}$ of a non-entangled state at finite temperature is a thermal mixture of products of local (reduced) density matrices $\rho_{A/B}$.
\begin{equation}
\label{eq:rhoAB}
    \rho_{AB} = \sum_{\lambda} p_\lambda \,\rho_{A}^\lambda \otimes \rho_{B}^\lambda.
\end{equation}
Measures of bipartite entanglement, specifically, the logarithmic negativity $N$ defined in Section {\it Methods}, detect whether the actual physical density matrix can be decomposed into such a separable product, with a non-zero value proving entanglement \cite{Horodecki_1996,Plenio_2005_logN,Vidal_2002_logN}.

However, since Alice and Bob perform local measurements on two spatially separated partitions (in this case lattice sites $i,\,j$), the measurements are subject to the underlying local symmetries of the system \cite{Schuch_2004_SSR}. In particular, local fermionic parity and, for the Fermi–Hubbard model, also local charge \cite{local_qinfo_Hubbard}, cannot be altered by local measurement protocols. These restrictions, the superselection rules (SSR), have profound implications for the entanglement structure of fermionic systems: only the fraction of the entanglement that is compatible with these SSR is operationally accessible for quantum information processing  \cite{Banhuls2007,Szalay2021}, making it distinct from the remaining non-accessible part 
Accessing entanglement between different superselection sectors would, among others, require operations that conflict with relativistic causality \cite{Friis_SSR_required_by_relativity_2016,Ding_2024_SSR}. 

Operationally, SSR can be enforced directly at the level of the density matrix \cite{Amico_entanglement,Ding_2024_SSR}. As illustrated in Fig.~\ref{Fig:main} (c), this amounts to projecting out matrix elements that connect distinct local superselection sectors and thus retaining only the SSR-allowed blocks (black and coloured entries), all of which are experimentally accessible in Fermi-Hubbard quantum simulators with spin and charge resolution. 
The only required off-diagonal component, a typical exchange coupling mediated via two hopping processes, can be reformulated using SU(2) symmetry as follows:
\begin{equation}
\begin{split}
\rho_{8,9} &= \big\langle \hat S^+_j \hat S^-_i \big\rangle = \frac{1}{2} \big\langle \hat S^+_j \hat S^-_i + \hat S^-_j \hat S^+_i \big\rangle 
= \big\langle \hat S^x_i \hat S^x_j + \hat S^y_i \hat S^y_j\big\rangle = 2 \big\langle \hat S^z_i \hat S^z_j \big\rangle.
\end{split}
\end{equation}
Here, $\hat S^{\eta}_i$
is the spin operator on site $i$ in direction $\eta=x,y,z$ respectively
flipping it for $\eta=\pm$.
This allows us to measure it in experiment, see Section {\it Methods}. The second  required (diagonal) density matrix element for the SSR negativity is 
\begin{equation}
    \rho_{7,7} =\big\langle (1-\hat n_{i\uparrow})\hat n_{i\downarrow}(1-\hat n_{j\uparrow})\hat n_{j\downarrow} \big\rangle ,
\end{equation}
expressed in terms of the occupation number operators $\hat n_{i\sigma}$ for a fermion on site $i$ with spin $\sigma$. 
By SU(2) symmetry,  $\rho_{10,10}= \rho_{7,7}$ and
$\rho_{9,8}=\rho_{8,9}$. 
The condition for SSR entanglement is that the red 2×2 block in Fig.~\ref{Fig:main}(c) develops a negative eigenvalue $\lambda$ under the partial transpose: $\lambda_{\pm} = \rho_{7,7} \pm\rho_{8,9}<0$. Since $\rho_{7,7}\geq 0$ for any thermal state, this is equivalent to $|\rho_{8,9}|>\rho_{7,7}$.
In the strong interaction (Heisenberg) limit at half filling, this criterion becomes equivalent to requiring the concurrence of two spins to be positive \cite{Amico_entanglement,Subrahmanyam_2004}.

Beyond formal consistency, SSR-constrained entanglement captures key features of strongly correlated electron systems, exhibiting a characteristic large value in the paramagnetic Mott insulating phase \cite{Bellomia2023} and for bad metals found in its proximity \cite{BellomiaPhD,Bellomia2026}.

We note that even a non-entangled density matrix can contain classical correlations between two bipartitions. The total amount of correlations, that is, classical and quantum correlations, including entanglement, can be quantified by the mutual Information $I$ \cite{Amico_entanglement,Henderson_2001}, which we also define in the Section {\it Methods} and explore in this work.

\section*{Results: Entanglement in the Fermi-Hubbard model}
In Fig.~\ref{Fig:main}(a), we present the nearest-neighbour SSR negativity $N_{NN}=N_{d=1}$, showing experimental data (diamonds) together with the D$\Gamma$A results (colour background) on a common colour scale as a function of temperature $T$ and electron doping $\delta$. Here, $T$ is measured in units of the hopping $t$, $\delta$ is relative to half-filling (negative $\delta$ corresponds to hole doping), and interactions are set to $U=6.5t$. The pseudogap regime is identified by a suppression of antinodal spectral weight in the analytically continued D$\Gamma$A spectral function $A(\mathbf{k},\omega)$ (see Supplemental Information S.1). The thus identified (solid) crossover line $T^*(\delta)$ in Fig.~\ref{Fig:main}(a) is --- within the uncertainties of a crossover line --- consistent with the doping-dependent temperature scale $\Theta(\delta)$ that drives the growth of magnetic correlations at the onset of the pseudogap reported in \cite{Chalopin_PG_2026} (see also \textit{Methods}).

Within the pseudogap regime, the SSR negativity between nearest neighbors  $N_{\rm NN}$ becomes finite (red or yellow color in Fig.~\ref{Fig:main}(a)). This establishes the presence of entanglement between nearest-neighbour spin degrees of freedom that are at the origin of the SSR restricted entanglement. The restriction to spin degrees of freedom can be understood when
inspecting the relevant matrix elements for entanglement (negativity $N_{\rm NN}>0$)  in Fig.~\ref{Fig:main}(c). These are the red elements that form an effective two-spin (two-qubit) subsector. In this setting, a vanishing negativity (black region) rigorously implies separability of the spin state. Therefore, our result demonstrates the absence of spin-singlet entanglement outside the pseudogap regime. Note that the largest experimentally observed value $N_{NN} = 0.0087(14)$ inside the pseudogap regime is clearly distinct from zero \footnote{Here, the estimated uncertainty is obtained from Gaussian error propagation of the bootstrap uncertainties of the relevant density matrix elements. Owing to the particle–hole symmetry of the model, theoretical data was computed only in the hole-doped regime (negative $\delta$), the exact data points are shown in Supplementary Information S.5}, demonstrating the presence of entanglement at the onset of the pseudogap regime.

To further our understanding, let us note that the negativity becomes finite precisely when the probability of detecting a spin singlet exceeds the joint probability of detecting any of the three spin-triplet states $p_{\rm singlet}/p_{\rm triplet} \geq 1$ (for a precise definition see Supplemental Information S.2). While both the spin-singlet and the $S^z=0$ triplet state are, in principle, entangled, the SU(2)-symmetric Hubbard model enforces equal weights for all three triplet components ($S^z=\pm1,0$). Their thermal mixture forms a separable state and therefore does not contribute to entanglement. If $p_{\rm singlet}/p_{\rm triplet} < 1$, the reduced density matrix can be described by an unentangled mixed state of the form Eq.~(\ref{eq:rhoAB}).

\begin{figure}[H]
    \centering
    \includegraphics[width=1.0\linewidth]{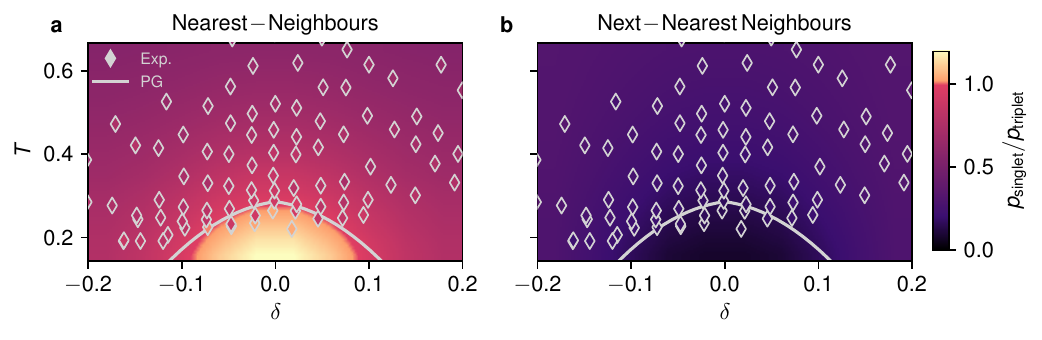}
    \caption{{\bf Ratio of spin-singlet to spin-triplet contributions.} SSR entanglement emerges for a ratio $p_{\rm singlet}/p_{\rm triplet} \geq 1$, shown as a function of $T$ and $\delta$ for (a) nearest-neighbour lattice sites and (b) next-nearest neighbours.  In (b),  $p_{\rm singlet}/p_{\rm triplet}$  even decreases in the pseudogap regime, i.e. the system moves further away from the entanglement threshold for next-nearest neighbor spins.}
    \label{fig:singlet_vs_triplet}
\end{figure}

Figure~\ref{fig:singlet_vs_triplet}\,(a) shows the singlet-to-triplet fraction in the experiment (diamonds) and for the theory (colour background). For nearest neighbours, this ratio increases upon approaching the pseudogap regime and exceeds the entanglement threshold just at its onset, consistent with the emergence of finite negativity.

In contrast, at larger distances, illustrated for next-nearest neighbours in Fig.~\ref{fig:singlet_vs_triplet}\,(b), the singlet-to-triplet ratio even decreases toward the pseudogap regime (note the colour change from purple to black) and remains well below the entanglement threshold $p_{\rm singlet}/p_{\rm triplet} < 1$ throughout. No SSR entanglement is therefore detected beyond nearest neighbours. Indeed, both the numerical background and the experimental diamonds in Fig.~\ref{fig:singlet_vs_triplet}\,(b) yield $p_{\rm singlet}/p_{\rm triplet} < 0.5$ at next-nearest neighbours, placing this conclusion well outside the numerical and experimental error bars, respectively. A detailed analysis of its robustness against experimental uncertainties is given in Supplementary Information~S.5.

Combining this behaviour with the structure of the SSR-restricted density matrix shown in Fig.~\ref{Fig:main}(c), we conclude that the observed entanglement is of spin-singlet character and confined to nearest neighbours only, as well as to the emergent pseudogap regime. At larger distances and outside the pseudogap regime, residual non-SSR entanglement arising from electronic hopping processes may persist, measured by the full fermionic negativity (see Supplemental Information S.4). However, this contribution lies outside the SSR-accessible sector and is therefore neither experimentally observable nor characteristic of the pseudogap regime. Such non-SSR ``hopping'' entanglement is even present in the metallic regime of the non-interacting model \cite{Bippus_2sRDM_entanglement_2026}.
 Connecting to the large $U$ limit, we note that also the antiferromagnetic Heisenberg model on the square lattice has spin entanglement between nearest-neighbors only \cite{Subrahmanyam_2004} with a much larger $p_{\rm singlet}/p_{\rm triplet}$ imbalance.  Finally, the observed spin-singlet entanglement gives a microscopic explanation for the increase of the quantum Fisher information in the pseudogap regime of cuprates \cite{Bippus_2025_pseudogap_QFI}.

\begin{figure}[H]
    \centering
    \includegraphics[width=1.0\linewidth]{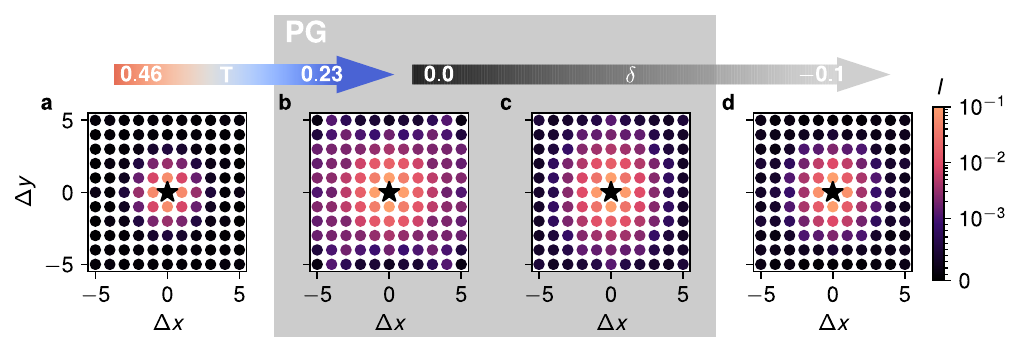}
   \caption{{\bf Experimental SSR mutual information.} The mutual information $I$ between two sites is shown as a function of relative displacement ($\Delta y$,$\Delta x$). It measures correlations and shows that these are present beyond nearest neighbours. From left to right, we first enter the pseudogap [(b), (c)] regime by lowering $T$ and then exit it again in (d) by increasing doping (cf.\ arrows).}
    \label{fig:I}
\end{figure}

Finally, Fig.~\ref{fig:I} demonstrates that, beyond the nearest-neighbour spin-singlet entanglement, longer-ranged classical correlations persist. Panels (a)–(d) display the spatial structure of the experimentally measured SSR mutual information $I$ between a central reference site (star) and its surrounding lattice sites for different temperatures $T$ and electron dopings $\delta$. This mutual information measures classical and quantum correlations from the same SSR-projected two-site density matrix \footnote{That is, all 18 black and colored elements in Fig.~\ref{Fig:main}\,(c).} for which the negativity is an entanglement measure. The mutual information in Fig.~\ref{fig:I} is short-ranged in the pseudogap regime but clearly goes beyond nearest neighbours. This is in contrast to SSR entanglement, which is zero except for nearest neighbors. The mutual information is actually rather similar --- qualitatively --- to the conventional (and specific) magnetic correlation function, see Supplemental Figure S.3 and \cite{Chalopin_PG_2026}, with a magnetic correlation length of about 2.5 lattice sites.

\section*{Discussion}

We observe, both numerically and experimentally, that spin-singlet entanglement in the Fermi-Hubbard model is directly connected to the onset of the pseudogap regime. In contrast to conventional correlations, this operationally accessible quantum entanglement is restricted to nearest neighbours only and is zero already for next-nearest neighbours and beyond. This is highly relevant for cuprate superconductors \cite{Gull2015_hubbard_model_for_cuprates}, where the pseudogap is evidenced in, among others, angle-resolved photoemission experiments \cite{Damacelli2003} and for which the Fermi-Hubbard model is paradigmatic. 
Let us note that for cuprates, it is established that the next-nearest-neighbor hopping $t'$ is crucial 
in accounting for the properties of the superconducting phase \cite{Pavarini2001,xu:2024,Roth2025}. Nevertheless, for realistic (i.e., small) values of $t'$ we expect qualitatively the same behavior as in our study with $t'= 0$.
Our work makes a sharp, falsifiable prediction connecting the pseudogap to the spin entanglement: increasing the next-nearest-neighbour hopping $|t'|$ must shift the negativity-onset line in lockstep with the pseudogap-onset line.

Our finding has direct implications for the modelling of the pseudogap phase and, by extension, possibly for mechanisms underlying high-temperature superconductivity. Without quantum mechanical superpositions the off-diagonal coherence $\rho_{8,9}$ can never become non-zero, which is required for a finite negativity. Moreover, a purely classical thermal mixture of singlet and triplet states can never exceed the entanglement threshold $p_{\rm singlet}/p_{\rm triplet}>1$. Hence, theories of the pseudogap that rely entirely on classical spin fluctuations are disfavoured by our data. A quantum component that becomes distinctly active at short distances in the pseudogap phase is required. An example of a theory that builds upon quantum spin entanglement between nearest neighbors is Anderson's resonating valence bond state \cite{Anderson1987}. Its resonating character cannot, however, be detected from the static reduced density matrix that we measured and calculated. Other theories of the pseudogap that develop nearest-neighbour singlet structure are also compatible with our data. 
We note that computations based on cluster extensions of DMFT clearly demonstrate the dominance of local singlet configurations in the pseudogap regime \cite{Haule2007,Ferrero_epl_2009,Ferrero_prb_2009,Bellomia2026}, see 
also \cite{Punk2015,Yu2024}. Testing this microscopic picture in further candidate models for the pseudogap, for example, the recently proposed ancilla wave functions~\cite{Muller_2025}, fluctuating stripes in geometric orthogonal metals \cite{schlomer_2025}, and fractionalized Fermi-liquids \cite{Bonetti_2026}, would provide valuable milestones towards a cohesive understanding of this enigmatic regime. 

To this end, our work provides the necessary tools to test any theory for two-site entanglement, as long as static spin-dependent density-density correlators between two sites can be calculated. Furthermore, our SSR  spin-singlet entanglement has the advantage of being a simple entanglement measure that can be directly determined in cold atom quantum simulators, also for general lattice models.

Let us finally mention that entanglement is also to be expected at and around quantum critical points \cite{hauke_measuring_2016}. However, for such a mechanism, the entanglement domes should be centered around the antiferromagnetic quantum phase transition around the critical doping of $\delta\sim \pm 0.1$. Our results show that entanglement already sets in (i) at higher 
temperatures, is (ii) centered around half filling, and is (iii) closely connected with the onset of the pseudogap. Our work does not, however, exclude additional entanglement appearing at lower temperatures from quantum criticality, superconducting, or stripe fluctuations. For now, it remains to be seen if such entanglement is short or long range and if it is of the operationally accessible kind.

\textbf{Outlook.}
Our work provides a first quantifiable test of the spatial structure of entanglement in the pseudogap: the SSR-restricted nearest-neighbour negativity in the two-dimensional Fermi-Hubbard model. 
Lowering the temperature into the regime where superconducting and
stripe correlations grow would test whether the entanglement there extends
beyond nearest neighbours.
Theoretically, compression algorithms \cite{Rohshap_2025_PA_QTT} and fundamental progress on self-consistent methods \cite{essl_2026, Lihm_2026} may allow for future D$\Gamma$A computations of the strongly correlated pseudogap beyond the ladder approximation.  
Experimentally, for models without SU(2) symmetry an interferometric or Ramsey readout can be employed to measure the
off-diagonal density matrix element $\rho_{8,9}$. In solid-state experiments it is $\rho_{7,7}$ that is the more challenging quantity, since accessing it requires up to fourth-order correlators, i.e., non-linear response measurements.


\section*{Acknowledgments}
We thank  A. Amaricci, F. Assaad,  A. Bohrdt, M. Capone, F. Grusdt, E. Jacob, C. Mejuto-Zaera, C. Schilling, and M. Ulybyshev  for fruitful discussions. This work has been supported in part by the Austrian Science Funds (FWF) through the FWF Spezialforschungsbereich (SFB) QM\&S project DOI 10.55776/F86, FWF project DOI 10.55776/V1018, and the European Research Council (ERC) through ERC-2024-ADG RealSuper project DOI 10.3030/101201037.
Calculations have been done on the Austrian Scientific Cluster (ASC).
This work was further supported by the Max Planck Society (MPG), the Horizon Europe program HORIZON-CL4-2022 QUANTUM-02-SGA (project 101113690, PASQuans2.1), the German Federal Ministry of Research, Technology and Space (BMFTR grant agreement 13N15890, FermiQP), and Germany’s Excellence Strategy (EXC-2111-390814868).
GB acknowledges further support by the Italian Ministry of University and Research
(MUR) via the PRIN 2022 (Prot.\,20228YCYY7) program.
The Flatiron Institute is a division of the Simons Foundation.
This project is funded in part by the European Union. Views and opinions expressed are however those of the author(s) only and do not necessarily reflect those of the European Union or the European Research Council Executive Agency. Neither the European Union nor the granting authority can be held responsible for them.

\section*{DATA AVAILABILITY}
The data that support the findings of this article are openly available \cite{}.

\section*{Competing interests}
The authors declare that they have no competing interests.

\bibliography{references_fixed.bib}


\section*{Methods}

Both experiment and theory simulate the 2D Fermi-Hubbard model
\begin{equation}
\begin{split}
    H = & -t\sum_{\langle ij\rangle, \sigma } \left( \hat c^{\dagger}_{i\sigma}\hat c_{j\sigma} + \text{h.c.} \right) \\
    & \quad \quad \quad + U \sum_{i} \hat n_{i\uparrow}\hat n_{i\downarrow} -\mu \sum_{i,\sigma} \hat n_{i,\sigma}
\label{eq:FHM}
\end{split}
\end{equation}
with only nearest-neighbour hoppings $t$ at Coulomb repulsion $U=6.5 t$ and chemical potential $\mu$ fixed to a given electron doping $\delta$. Here, $\hat c^{(\dagger)}_{i\sigma}$ are fermionic annihilation (creation) operators on lattice site $i$  with spin-$\frac{1}{2}$ $\sigma = \{\uparrow,\downarrow\}$ in $z$ direction; $\hat n_{i\sigma}=\hat c^{\dagger}_{i\sigma}\hat c_{i\sigma}$. We set $\hbar=k_{\mathrm{B}}=1$, and measure energies in units of $t=1$.

\subsection*{Experiment}

The experimental protocol can be found in detail in \cite{Chalopin_PG_2026} and in references therein. We recall here the main aspects.

We load a spin-balanced ultracold gas of $^{6}$Li atoms in a 2D optical lattice \cite{Bloch_QSIM_revieew}, naturally implementing the 2D Fermi-Hubbard model given in Eq.~\eqref{eq:FHM}. The optical potential is engineered to realise a homogeneous disk-shape system of about $145$ lattice sites, surrounded by a low-density region whose chemical potential is adjusted to control the doping of the central region. The temperature is tuned by holding the atoms in the lattice for a variable time, naturally causing heating \cite{mazurenko:2017a}.
Detection is performed using a spin-resolved quantum gas microscope \cite{gross:2021} whose technical description can be found in Refs. \cite{bourgund:2025, Chalopin_optical_superalattices_2025}.
Thermometry is achieved by comparing the measured spin-spin correlations to determinant Quantum Monte Carlo simulations performed at various doping and temperatures \cite{blankenbecler:1981, varney:2009}.

The range of temperatures and dopings explored by the experiment is illustrated by the diamonds in Figs.~\ref{Fig:main}, \ref{fig:singlet_vs_triplet}. The pseudogap regime, in particular, is reached in the coldest datasets ($T/t \lesssim 0.3$) close to half-filling ($|\delta| \lesssim 10\%$). In this regime, an exponential increase of the equal-time spin structure factor is observed as the temperature decreases, driven by a doping-dependent temperature scale $\Theta(\delta)$ that compares quantitatively to the pseudogap temperature $T^*$ \cite{Chalopin_PG_2026}.
Additional details regarding the measurement of $\Theta(\delta)$ can be found in the Supplemental Information S.1.

Quantum gas microscopy allows us to reconstruct all number-operator correlations at arbitrary order of the form $\langle \hat n_i \hat n_j \cdots \hat n_k \rangle$ \cite{chalopin2025connectedcorrelationscoldatom}.
Spin resolution furthermore allows to measure spin-spin correlations, \emph{i.e.} correlators of the form $\langle \hat S^z_i \hat S^z_j \rangle$, where $\hat S^z_i = (\hat n_{i,\uparrow} - \hat n_{i,\downarrow})/2$. From these correlators, all the elements of the two-site density matrix that remain after application of the SSR (see Fig.~\ref{Fig:main}c) are accessed experimentally, cf.\ Supplemental Information S.3. In practice, the expectation value $\langle \cdot \rangle$ is estimated by averaging over hundreds to thousands of snapshots acquired in the same experimental conditions. Note that some, but not all, of these correlators could be reused from \cite{Chalopin_PG_2026}.

\subsection*{Theory}
For theory, we use ladder dynamical vertex approximation (D$\Gamma$A) with $\lambda$ correction, a diagrammatic non-local extension to dynamical mean field theory (DMFT) \cite{rohringer_diagrammatic_2018,Toschi2007}. It includes all local DMFT correlations non-perturbatively but also non-local spin and charge fluctuations through ladder diagrams in terms of a local two-particle irreducible vertex $\Gamma$. We use {w2dynamics} \cite{Wallerberger_CompPhysComm_2019_w2dynamics} for calculating $\Gamma$ at DMFT convergence and subsequently the DGApy code  \cite{worm_numerical_2023} for the ladder diagrams.

As a postprocessing step, the two-site reduced density matrix is computed from two- and four-point Green's functions according to \cite{Roosz_2024_2sRDM}. Computations were performed on a $16\times 16$ lattice with periodic boundary conditions.

\subsection*{Entanglement measures and SSR}
\begin{figure}[H]
    \centering
    \includegraphics[width=1.0\linewidth]{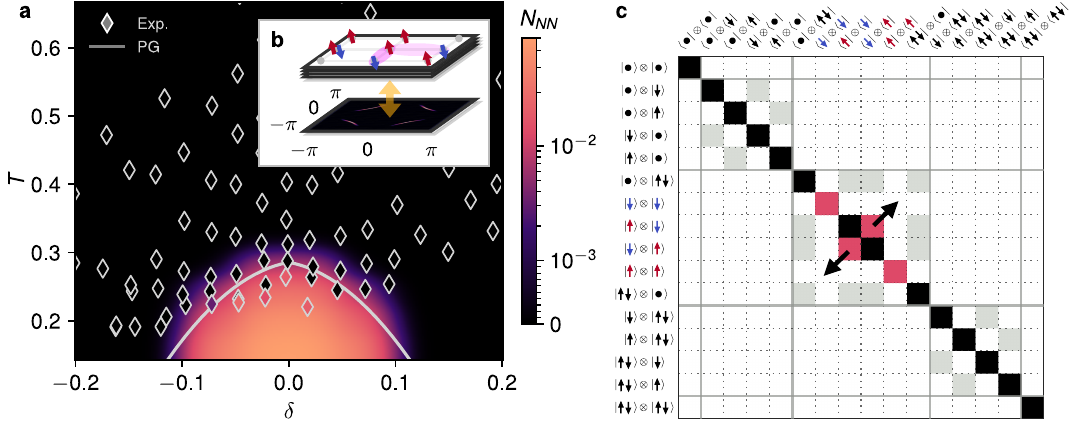}
    \caption{{\bf Entanglement in the Fermi-Hubbard model.} (a) Quantum entanglement between nearest-neighbour lattice sites as a function of temperature $T$ and doping $\delta$, measured through the SSR negativity $N_{\rm NN}$ and comparing quantum simulation experiments (diamonds) to D$\Gamma$A calculations (background). (b) Schematics of spin-singlet configurations that contribute to this entanglement (top) within the pseudogap regime (bottom, solid PG line in (a)). (c) Full two-site density matrix. Remaining elements after application of the SSR are marked in black and red (grey elements are finite before being projected out by SSR). Only the red elements contribute to the SSR negativity. 
    Black arrows indicate how these off-diagonal components are transformed by acting with the partial transpose on the SSR density matrix. Physically, the red off-diagonal component corresponds to spin fluctuations, as highlighted by the coloured spin labels (red for up, blue for down) in the basis labels at the top and left.
    }
    \label{Fig:main}
\end{figure}

Given a bipartite density matrix $\rho_{ij}$ between lattice sites ($i,\, j$), in the Fock basis \linebreak $|n_{i\uparrow},n_{i\downarrow}; n_{j\uparrow},n_{j\downarrow}\rangle = |n_{i\uparrow},n_{i\downarrow}\rangle \otimes |n_{j\uparrow},n_{j\downarrow}\rangle$, the charge SSR (which also enforces the parity restriction) is formally applied by a projection $P_{\mathcal{N}_{i/j}}$ onto the subspace of locally conserved electron number $\mathcal{N}_{i/j}$ 
\begin{equation}
    \rho_{ij}^{\mathrm{NSSR}} = \sum_{\mathcal{N}_i,\mathcal{N}_j} P_{\mathcal{N}_i} \otimes P_{\mathcal{N}_j} \rho_{ij} P_{\mathcal{N}_i} \otimes \mathcal{P}_{\mathcal{N}_j}.
\end{equation}
If such a bipartite density matrix cannot be expressed in a thermal mixture of product states
\begin{equation}
    \rho_{ij}^{\mathrm{NSSR}} = \sum_{\lambda} p_\lambda \rho_{i}^{\lambda} \otimes \rho_{j}^{\lambda},
\end{equation}
it is entangled
. According to the Peres-Horodecki criterion, this is detected by the appearance of negative eigenvalues under the partial transpose (see Supplemental Information S.4 for a definition of the partial transpose, which here transforms the SSR-restricted reduced density matrix as indicated by the black arrows in Fig.\ref{Fig:main}\,(c)) \cite{Peres_1996,Horodecki_1996,Amico_entanglement}. The amount of entanglement is measured by the (logarithmic-)negativity \cite{Plenio_2005_logN,Vidal_2002_logN}
\begin{equation}\label{eq_negativity}
    N = \log_2\left( \| \rho_{ij}^{T_j} \|_{\textrm{tr}} \right),
\end{equation}
with trace norm $\| \rho \|_{\textrm{tr}} = \textrm{tr}\sqrt{\rho \rho^{\dagger}}$ \cite{Vidal_2002_logN}. Due to the density matrix being positive semi-definite, only a small subspace can contribute to the SSR entanglement (coloured elements in Fig.~\ref{Fig:main} (c). This subspace is small enough for the SSR negativity to be a faithful measure i.e.: $N=0$ corresponds to no entanglement. Moreover, this subspace constitutes spin-singlet entanglement, which emerges when the singlet state dominates over the combined triplet states.

We further consider the SSR mutual information $I$ \cite{Ding_quant_chem_2021,Amico_entanglement}, based on the relative von Neumann entropy
\begin{equation}
    I = {\mathcal S}_{i}+{\mathcal S}_{j} - {\mathcal S}_{ij}.
\end{equation}
It measures the difference between a hypothetical uncorrelated system, expressed by the local von Neumann entropy ${\mathcal S}_{i} = -\textrm{Tr}[\rho_{i}\ln\rho_{i}]$ of individual sites $i$ and $j$, and the actual physical von Neumann entropy of both sites ${\mathcal S}_{ij}= -\textrm{Tr}[\rho_{ij}\ln\rho_{ij}]$. The mutual information measures all correlations
encoded in the remaining off-diagonal components of  $\rho_{i,j}$, including entanglement, other quantum mechanical, and classical correlations \cite{Henderson_2001}.

\newpage

\end{document}



\title{Supplemental Information: Emergence of spin entanglement with the pseudogap onset in the Fermi-Hubbard model}

\author{Frederic Bippus \orcidlink{0009-0006-4316-6547}}
\affiliation{Institute of Solid State Physics, TU Wien, 1040 Vienna, Austria}

\author{Thomas Chalopin \orcidlink{0000-0001-7633-0442}}
\affiliation{Université Paris-Saclay, Institut d’Optique Graduate School, CNRS, Laboratoire Charles Fabry, Palaiseau 91127, France}
\affiliation{Max-Planck-Institut für Quantenoptik, Garching 85748, Germany}
\affiliation{Munich Center for Quantum Science and Technology, Munich 80799, Germany}

\author{Gabriele Bellomia~\orcidlink{0000-0003-1576-3388}}
\affiliation{Institute of Solid State Physics, TU Wien, 1040 Vienna, Austria}
\affiliation{Scuola Internazionale Superiore di Studi Avanzati (SISSA), 34136 Trieste, Italy}

\author{Gerg\H o Ro\'osz \orcidlink{0000-0002-9541-119X}} 
\affiliation{Institute of Solid State Physics, TU Wien, 1040 Vienna, Austria}

\author{Titus Franz \orcidlink{0009-0007-8363-2896}}
\affiliation{Max-Planck-Institut für Quantenoptik, Garching 85748, Germany}
\affiliation{Munich Center for Quantum Science and Technology, Munich 80799, Germany}

\author{Antoine Georges \orcidlink{0000-0001-9479-9682}}
\affiliation{Collège de France, Paris Sciences et Lettres University, Paris 75005, France}
\affiliation{Center for Computational Quantum Physics, Flatiron Institute, New York, NY 10010}
\affiliation{École Polytechnique, Centre de Physique Théorique, CNRS, Paris F-91128, Palaiseau, France}
\affiliation{Department of Quantum Matter Physics, Université de Genève, Genève CH-1211, Suisse}

\author{Anna Kauch \orcidlink{0000-0002-7669-0090}}
\affiliation{Institute of Solid State Physics, TU Wien, 1040 Vienna, Austria}

\author{Immanuel Bloch \orcidlink{0000-0002-0679-4759}}
\affiliation{Max-Planck-Institut für Quantenoptik, Garching 85748, Germany}
\affiliation{Fakultät für Physik, Ludwig-Maximilians-Universität, Munich 80799, Germany}
\affiliation{Munich Center for Quantum Science and Technology,
Munich 80799, Germany}

\author{Karsten Held \orcidlink{0000-0001-5984-8549}}
\affiliation{Institute of Solid State Physics, TU Wien, 1040 Vienna, Austria}
    
\maketitle

{\bf In this supplemental information, we discuss{:} (1) the detection of the pseudogap regime; (2) the details behind the interpretation of the spin singlet as the driver of entanglement; (3) the two-site reduced density matrix and how to obtain it from the experiment and the dynamical vertex approximation; (4) the full fermionic negativity that captures all entanglement in the system, including the operationally inaccessible part beyond SSR; (5) experimental uncertainties and the absence of SSR entanglement beyond nearest neighbors.}

\section{Detection of the pseudogap regime}
Since the pseudogap (PG) regime has a cross-over to the metallic regime, determining its boundaries is not a well-defined task. Within the ladder dynamical vertex approximation (D\textGamma A), Matsubara frequency $i\omega_n$ and momentum $\mathbf{k}$ resolved one- and two-particle correlators can be obtained. Therefore, the momentum resolved spectral function $A(\omega,\mathbf{k})$ as the most direct measure of the pseudogap can be applied. The pseudogap regime is characterized by a partially opened gap around the antinode (AN) at $\mathbf{k} \approx (0,\pi)$ while spectral weight at the node (N) $\mathbf{k} \approx (\frac{\pi}{2},\frac{\pi}{2})$ remains large. We obtain $A(\omega,\mathbf{k})$ from the analytically continued D\textGamma A self-energy $\Sigma$. The analytical continuation is performed with the maximum entropy method in its $\chi$-to-kink formulation \cite{Kaufmann_ana_cont_2023}. 

\begin{figure*}
    \centering
    \includegraphics[width=1.0\linewidth]{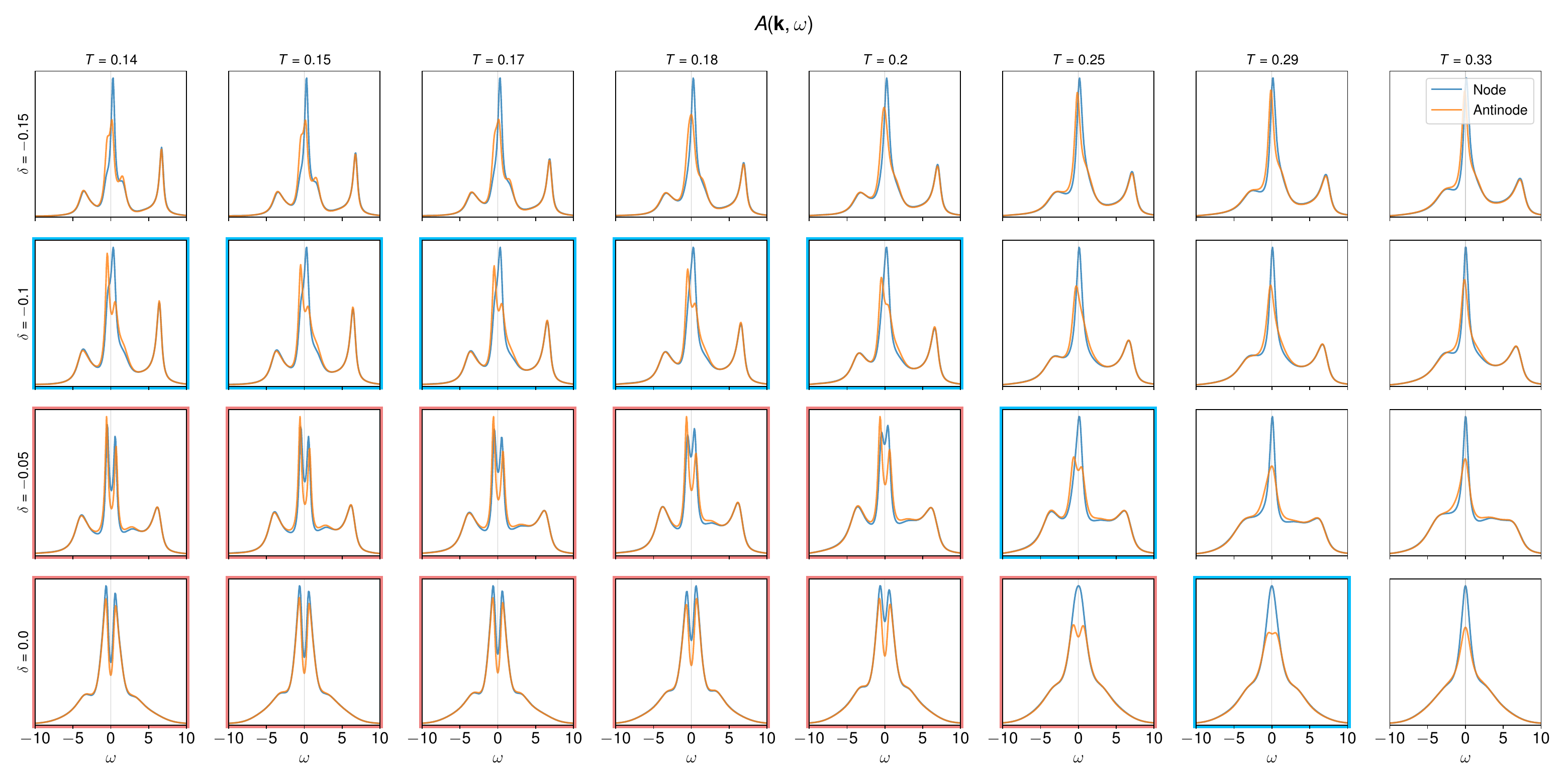}
    \caption{D\textGamma A spectral function $A(\mathbf{k_{\textrm{N/AN}}},\omega)$ at node and antinode obtained from analytical continuation with the maximum entropy method for different doping $\delta$ and temperatures $T$. A red borderline highlights spectra within the pseudogap regime, and a blue one those at its onset. }
    \label{fig:Spectral_functions}
\end{figure*}

In Fig.~\ref{fig:Spectral_functions} we show the resulting spectral function at the node and antinode. Results are consistent with a direct analytical continuation of the Green's function. We use a quadratic fit through the data-points highlighted in blue as the boundary for the PG in the main text, these elements are the first ones to show a pseudogap behaviour in the analytical continued data.

This direct detection of the PG in the theoretical D\textGamma A data is in good agreement with the experimental onset of the PG and previous theoretical minimally entangled typical thermal states simulations (METTS) results defined in Chalopin et al. \cite{Chalopin_PG_2026}.
There, the onset of the pseudogap is confirmed numerically by the maximum in the equal-time spin susceptibility at temperature $T^*$ (see also \cite{wietek:2021}), following the original description of the pseudogap in cuprates \cite{Johnston_downturn_in_chi_1989, alloul:1989}.
The experimental data furthermore reveal an exponential increase of the equal-time spin structure factor at $(\pi, \pi)$ as the temperature is reduced, following a doping-dependent temperature scale $\Theta(\delta)$ which compares quantitatively with $T^*$.
Note that, compared to $T^*$ and $\Theta(\delta)$ observed in \cite{Chalopin_PG_2026}, the D\textGamma A calculations show the onset of an antinodal gap at a slightly lower temperature.

\section{Spin singlet driven entanglement}
In \cite{Ding_SSR_entanglement_condition_2022}, a condition for the emergence of SSR entanglement is derived which, in a $\mathrm{SU}(2)$ symmetric system, can be written as \begin{equation}\label{ps_pt_condition}
    p_{\rm singlet}/p_{\rm triplet} \equiv \frac{\textrm{Tr}[P_{\rm singlet}\,\hat\rho_{ij}]}{\sum_{\alpha=1}^{3} \textrm{Tr}[P_{\rm triplet_{\alpha}}\,\hat\rho_{ij}]} > 1,
\end{equation}
with the projector $P_{|\psi\rangle}=|\psi\rangle\langle\psi|$ onto the singlet and three triplet states
\begin{align}
    |\psi_{\rm singlet}\rangle &= \frac{1}{\sqrt{2}}\left(|\uparrow,\downarrow\rangle - |\downarrow,\uparrow\rangle \right), \\
    |\psi_{\rm triplet_+}\rangle &= |\uparrow,\uparrow\rangle , \\
    |\psi_{\rm triplet_0}\rangle &= \frac{1}{\sqrt{2}}\left(|\uparrow,\downarrow\rangle + |\downarrow,\uparrow\rangle \right), \\
    |\psi_{\rm triplet_-}\rangle &= |\downarrow,\downarrow\rangle .
\end{align}
We derive the condition by noting that all diagonal elements of the density matrix are positive semidefinite, so the only two eigenvalues of the partial-transposed density matrix that can become negative are
\begin{equation}\label{cond_rho}
    \lambda_{\pm} = \rho_{7,7} \pm \rho_{8,9},
\end{equation}
which (since $\rho_{7,7} \geq 0$) translates to the condition that SSR entanglement is detected if
\begin{equation} \label{eq_condition_bare}
    \frac{|\rho_{8,9}|}{\rho_{7,7}} > 1.
\end{equation}
From $\mathrm{SU}(2)$ symmetry it follows that $\rho_{8,9} = 1/2 (\rho_{8,9} + \rho_{9,8})$, which can also be seen from the requirement of hermiticity of $\hat\rho$ and the fact that all density matrix elements are purely real \cite{Roosz_2024_2sRDM}. Using the operator relation
\begin{equation}
    \frac{1}{2} \left( \hat\rho_{8,9} + \hat\rho_{9,8} \right) = \frac{1}{2} \Big( |\uparrow,\downarrow\rangle\langle\downarrow,\uparrow| + |\downarrow,\uparrow\rangle\langle\uparrow,\downarrow| \Big) = \frac{1}{2}\Big( |\psi_{\rm triplet_0}\rangle\langle\psi_{\rm triplet_0}| - |\psi_{\rm singlet}\rangle\langle\psi_{\rm singlet}|\Big)
\end{equation}
we rewrite Eq.~\eqref{eq_condition_bare} in terms of the following expectation values
\begin{equation}\label{intermediate_condition}
    \frac{|\rho_{8,9}|}{\rho_{7,7}} = \frac{\Big|\textrm{Tr}\Big[\frac{1}{2}\Big( |\psi_{\rm triplet_0}\rangle\langle\psi_{\rm triplet_0}| - |\psi_{\rm singlet}\rangle\langle\psi_{\rm singlet}|\Big) \hat\rho_{ij}\Big]\Big|}{ \textrm{Tr}\Big[|\psi_{\rm triplet_0}\rangle\langle\psi_{\rm triplet_0}|\hat\rho_{ij}\Big]} > 1.
\end{equation}
Here we have used $p_{T_0}=\mathrm{Tr}[|\psi_{\rm triplet_0}\rangle\langle\psi_{\rm triplet_0}|\hat\rho_{ij}] = \rho_{7,7}$, an identity that follows from $\mathrm{SU}(2)$ symmetry ($p_{T_+}=p_{T_0}=p_{T_-}$). The condition Eq.~\eqref{intermediate_condition} is evidently fulfilled if the following holds
\begin{equation}
    p_{\rm singlet} > 3 p_{\rm triplet_0} = p_{\rm triplet}.
\end{equation}
Hence, we arrive at Eq.~\eqref{ps_pt_condition}.

At nearest-neighbour distance, this condition is fulfilled in the pseudogap regime while the ratio drops for lattice sites at larger distances, see Fig.~2 of the main paper. Nevertheless, we want to stress that the dominant eigenvector of the full D\textGamma A 2s-RDM is a superposition of the singlet and holon-doublon
\begin{equation}
    |\psi_{\textrm{dominant}}\rangle = a\underbrace{\Big(|0,\uparrow\downarrow\rangle + |\uparrow\downarrow,0\rangle \Big)}_{\textrm{holon-doublon}} +b\underbrace{\Big(|\uparrow,\downarrow\rangle - |\downarrow,\uparrow\rangle \Big)}_{\textrm{\rm singlet}},
\end{equation}
However, SSR distills the singlet entanglement that is usable for quantum information protocols from it. This is in good agreement with previous CDMFT results \cite{Georges_2009_orbital_selective_MIT, Tremblay_2011_normal_state_phase_diagramm} which have identified this state as dominant in the bad metal regime.

\section{Two-site reduced density matrix}
\begin{figure*}
    \centering
    \includegraphics[width=1.0\linewidth]{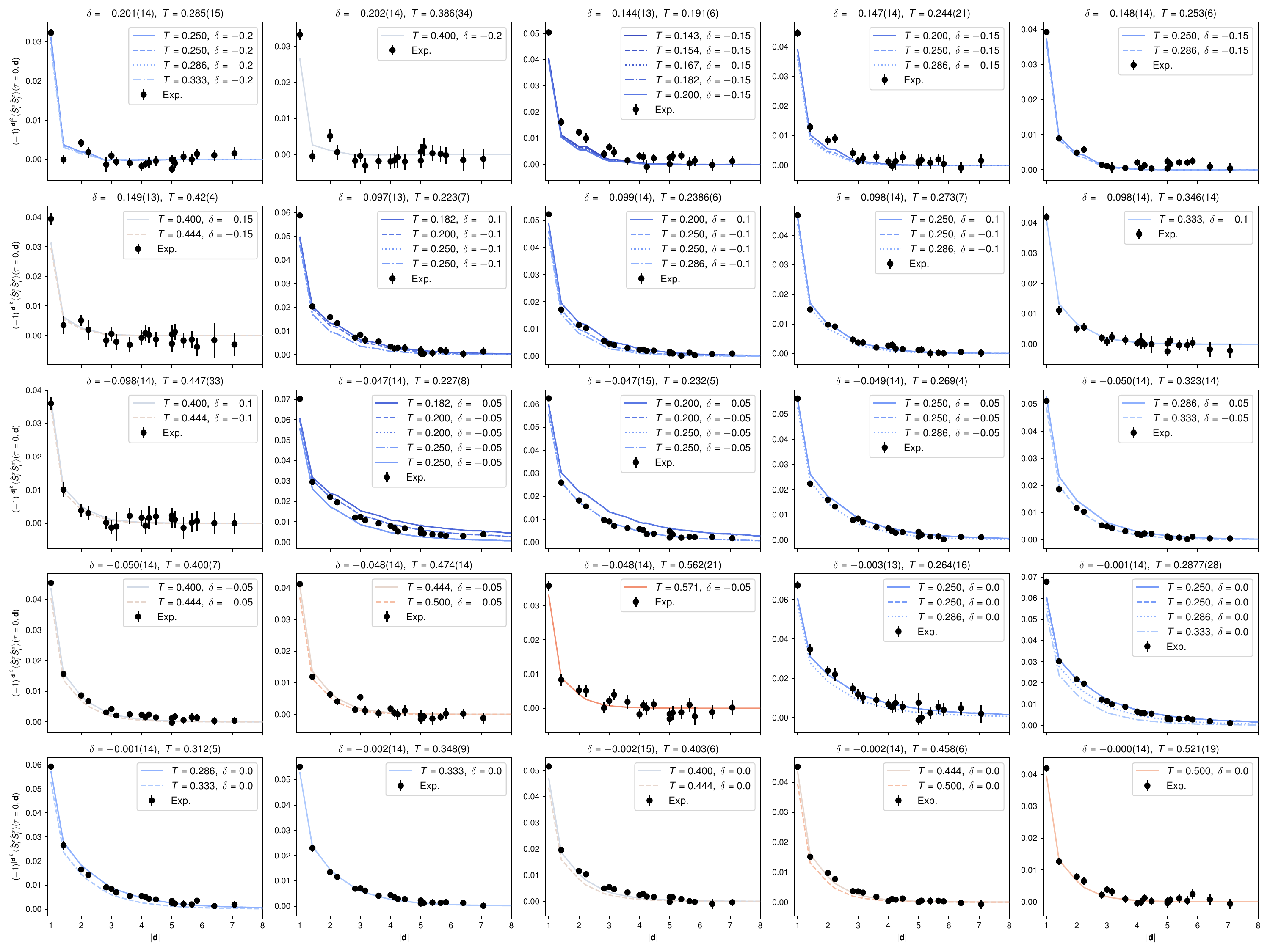}
    \caption{Comparison of D$\Gamma$A and experimental equal-time spin-spin correlator $(-1)^{\mathbf{d}^2}\langle \hat S^z_i \hat S^z_j \rangle(\tau=0,\mathbf{d})$ over distance $|\mathbf{d}|$ for comparable dopings $\delta$ and temperatures $T$, with the factor $(-1)^{\mathbf{d}^2}$ removing the antiferromagnetic sign. 
    \label{fig:susceptibility_comparison}}
\end{figure*}

Quantum simulator experiments offer access to all connected correlators that are based on combinations of number operators $\hat n_{i\sigma} = \hat c^{\dagger}_{i\sigma}\hat c_{i\sigma}$ \cite{chalopin2025connectedcorrelationscoldatom}. This is sufficient to reconstruct all SSR-allowed elements of the two-site reduced density matrix (see Fig.~1\,(c) of the main text):
\begin{subequations}
\begin{align}
	\rho_{1,1} &= \langle\, (1-\hat n_{i\uparrow})(1-\hat n_{i\downarrow})  (1-\hat n_{j\uparrow})(1-\hat n_{j\downarrow}) \, \rangle \\
	\rho_{2,2} &= \langle\,  (1-\hat n_{i\uparrow})(1-\hat n_{i\downarrow})(1-\hat n_{j\uparrow})\hat n_{j\downarrow} \, \rangle \\
	 \rho_{3,3} &= \langle\, (1-\hat n_{i\uparrow})(1-\hat n_{i\downarrow}) \hat n_{j\uparrow}(1-\hat n_{j\downarrow}) \, \rangle \\
	\rho_{4,4} &=\langle\,  (1-\hat n_{i\uparrow})\hat n_{i\downarrow} (1-\hat n_{j\uparrow})(1-\hat n_{j\downarrow}) \, \rangle \\
	 \rho_{5,5} &=\langle\,  \hat n_{i\uparrow}(1-\hat n_{i\downarrow})(1-\hat n_{j\uparrow})(1-\hat n_{j\downarrow}) \, \rangle \\
	 \rho_{6,6} &=\langle\,  (1-\hat n_{i\uparrow})(1-\hat n_{i\downarrow}) \hat n_{j\uparrow}\hat n_{j\downarrow} \, \rangle \\
	 \rho_{7,7} &=\langle\,  (1-\hat n_{i\uparrow})\hat n_{i\downarrow}(1-\hat n_{j\uparrow})\hat n_{j\downarrow} \, \rangle \\
	 \rho_{8,8} &=\langle\,  \hat n_{i\uparrow} (1-\hat n_{i\downarrow})(1-\hat n_{j\uparrow})\hat n_{j\downarrow} \, \rangle \\
	 \rho_{9,9} &=\langle\,  (1-\hat n_{i\uparrow}) \hat n_{i\downarrow}  \hat n_{j\uparrow}(1-\hat n_{j\downarrow}) \, \rangle \\
	 \rho_{10,10} &=\langle\,  \hat n_{i\uparrow}  (1-\hat n_{i\downarrow})  \hat n_{j\uparrow} (1-\hat n_{j\downarrow}) \, \rangle \\
	\rho_{11,11} &=\langle\,  \hat n_{i\uparrow}\hat n_{i\downarrow}(1-\hat n_{j\uparrow})(1-\hat n_{j\downarrow}) \, \rangle \\
	\rho_{12,12} &=\langle\,  (1-\hat n_{i\uparrow})\hat n_{i\downarrow} \hat n_{j\uparrow}\hat n_{j\downarrow} \, \rangle \\
	\rho_{13,13} &=\langle\,  \hat n_{i\uparrow}(1-\hat n_{i\downarrow}) \hat n_{j\uparrow}\hat n_{j\downarrow} \, \rangle \\
	\rho_{14,14} &=\langle\,  \hat n_{i\uparrow}\hat n_{i\downarrow} (1-\hat n_{j\uparrow})\hat n_{j\downarrow} \, \rangle \\
	\rho_{15,15} &=\langle\,  \hat n_{i\uparrow}\hat n_{i\downarrow} \hat n_{j\uparrow}(1-\hat n_{j\downarrow}) \, \rangle \\
	\rho_{16,16} &=\langle\,  \hat n_{i\uparrow}\hat n_{i\downarrow} \hat n_{j\uparrow}\hat n_{j\downarrow} \, \rangle.
\end{align}
Also the equal-time spin-spin correlator
\begin{equation} \label{eq_SS_correlator}
    \rho_{8,9} = 2\langle \, \hat S^z_i  \hat S^z_j \, \rangle
\end{equation}
\end{subequations}
can be measured from $n_{i\sigma}$, since
\begin{equation}
    \hat S^z_i = \frac{1}{2}\left(\hat n_{i\uparrow} - \hat n_{i\downarrow} \right)\, .
\end{equation}
The explicit expression of this matrix element follows from $\mathrm{SU}(2)$ symmetry \cite{Roosz_2024_2sRDM}.
In Fig.~\ref{fig:susceptibility_comparison} we show excellent agreement in a direct comparison of $\rho_{8,9}$ from experiment and D\textGamma A.

\begin{figure*}
    \centering
    \includegraphics[width=1.0\linewidth]{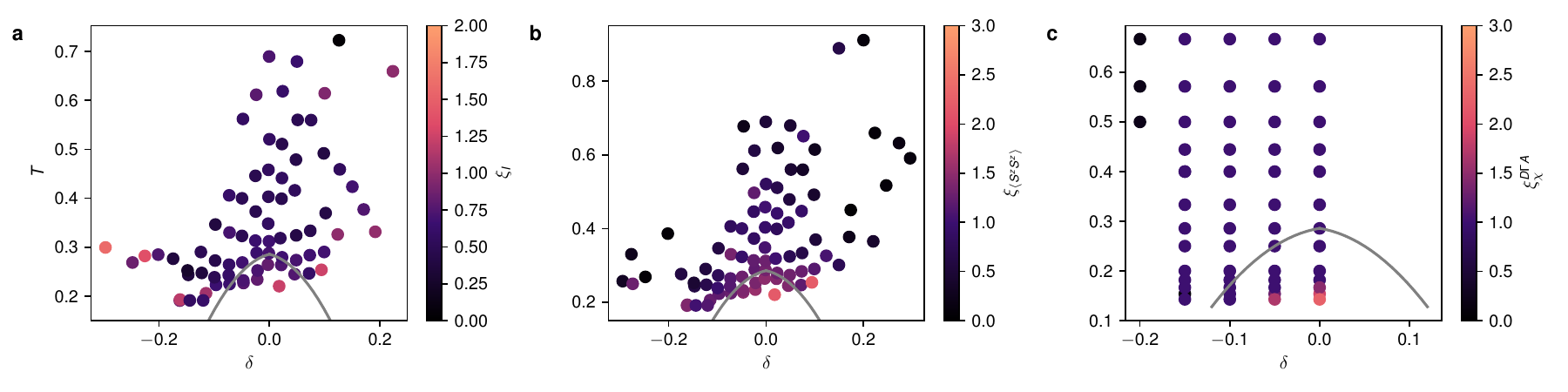}
    \caption{Correlation length $\xi$ over doping $\delta$ and temperature $T$ for (a) the mutual information $I$ from experiment; (b) the equal-time spin-spin correlator from experiment; (c) for the D$\Gamma$A spin susceptibility at the zeroth Matsubara frequency. 
    \label{fig:corr_length}}
\end{figure*}

We furthermore provide the correlation length $\xi$, $\langle \hat S^z_i \hat S^z_j \rangle \propto e^{-r/\xi}$ obtained from experimental equal-time spin-spin correlator, mutual information, and static Matsubara frequency D$\Gamma$A spin susceptibility in Fig.~\ref{fig:corr_length}. While the correlation length from the mutual information remains rather short $\sim 1.5$ sites, it shows a significant increase towards the pseudogap regime. The magnetic (spin) correlation length is slightly larger: up to $\sim 2.5$ sites.

Numerical methods can further access density matrix elements that break local charge conservation and also parity constraints \cite{Roosz_2024_2sRDM,Bippus_2sRDM_entanglement_2026,Bellomia2023}. Specifically the remaining off-diagonal components (shown in grey in Fig.~1\,(c) of the main text) can be recovered from 
\begin{subequations}
    \begin{align}
\rho_{2,4}&=\frac{1}{2} \langle ( \hat c^{\dagger}_{i\downarrow} \hat c_{j \downarrow}+\hat c^{\dagger}_{j\downarrow}\hat c_{i\downarrow})(1-\hat n_{i\uparrow})(1-\hat n_{j,\uparrow}) \rangle,\\
\rho_{3,5}&=\frac{1}{2} \langle( \hat c^{\dagger}_{i\uparrow}\hat c_{j\uparrow}+\hat c^{\dagger}_{j\uparrow}\hat c_{i\uparrow})(1-\hat n_{i\downarrow})(1-\hat n_{j,\downarrow}) \rangle,\\
\rho_{8,6}&=\frac{1}{2} \langle( \hat c^{\dagger}_{i\uparrow}\hat c_{j\uparrow}+\hat c^{\dagger}_{j\uparrow}\hat c_{i\uparrow})(1-\hat n_{i\downarrow})\hat n_{j,\downarrow} \rangle,\\
\rho_{9,6}&=\frac{-1}{2} \langle( \hat c^{\dagger}_{i\uparrow}\hat c_{j\uparrow}+\hat c^{\dagger}_{j\uparrow}\hat c_{i\uparrow})(1-\hat n_{i\uparrow}) \hat n_{j,\uparrow}  \rangle,\\
\rho_{11,6}&=\frac{1}{2} \langle(  \hat c^{\dagger}_{i\uparrow}\hat c^{\dagger}_{i \downarrow} \hat c_{j\downarrow}\hat c_{j\uparrow} 
 +  \hat c^{\dagger}_{j\uparrow} \hat c^{\dagger}_{j\downarrow} \hat c_{i \downarrow} \hat c_{i\uparrow}) \rangle,\\
 \rho_{8,11}&=\frac{1}{2} \langle( \hat c^{\dagger}_{j\downarrow}\hat c_{i,\downarrow}+\hat c^{\dagger}_{i\downarrow}\hat c_{j\downarrow})\hat n_{i\uparrow}(1-\hat n_{j,\uparrow}) \rangle,\\
 \rho_{9,11}&=\frac{-1}{2} \langle( \hat c^{\dagger}_{j\uparrow}\hat c_{i,\uparrow}+\hat c^{\dagger}_{i\uparrow}\hat c_{j\uparrow})\hat n_{i\downarrow}(1-\hat n_{j,\downarrow}) \rangle,\\
 \rho_{12,14}&=\frac{-1}{2} \langle( \hat c^{\dagger}_{j\uparrow}\hat c_{i\uparrow}+\hat c^{\dagger}_{i\uparrow}\hat c_{j\uparrow})\hat n_{i\downarrow}\hat n_{j,\downarrow} \rangle,\\
 \rho_{13,15}&=\frac{-1}{2} \langle( \hat c^{\dagger}_{j\downarrow}\hat c_{i\downarrow}+\hat c^{\dagger}_{i\downarrow}\hat c_{j\downarrow})\hat n_{i\uparrow}\hat n_{j\uparrow}\rangle .
\label{eq:non-diag}
\end{align}
\end{subequations}
This is in the Fock basis defined in \cref{tab:basis}.
\begin{table}[tb]
	\begin{center}
		\begin{tabular}{c |c c | c c| } 
             & \multicolumn{2}{c|}{$\psi_i$} & \multicolumn{2}{c|}{$\psi_j$} \\ \cline{2-5}
			 & i$\uparrow$ & i$\downarrow$  & j$\uparrow$  & j$\downarrow$  \\ \hline
		   $v_1$ & 0 & 0  & 0  & 0 \\ \hline
		   $v_2$ & 0 & 0  & 0  & 1 \\
		   $v_3$ & 0 & 0  & 1  & 0 \\
		   $v_4$ & 0 & 1  & 0  & 0 \\
		   $v_5$ & 1 & 0  & 0  & 0 \\ \hline
		   $v_6$ & 0 & 0  & 1  & 1 \\
		   $v_7$ & 0 & 1  & 0  & 1 \\
		   $v_8$ & 1 & 0  & 0  & 1 \\
		   $v_9$ &  0 & 1  & 1  & 0 \\
		   $v_{10}$ & 1 & 0  & 1  & 0 \\
		   $v_{11}$ & 1 & 1  & 0  & 0 \\ \hline
		   $v_{12}$ & 0 & 1  & 1  & 1 \\
		   $v_{13}$ & 1 & 0  & 1  & 1 \\
		   $v_{14}$ & 1 & 1  & 0  & 1 \\
		   $v_{15}$ & 1 & 1  & 1  & 0 \\ \hline
		   $v_{16}$ & 1 & 1  & 1  & 1 
		\end{tabular}
	\end{center}
		\caption{Occupation number basis of all two-site states $v$, expressed in terms of the 
        occupations of the two sites $i$ and $j$ and two spin species $\uparrow$ and $\downarrow$; $\psi_i$ and   $\psi_j$  denote the basis states for site $i$ and $j$ from which  the product state $v$ is made up.  Horizontal lines separate subspaces with different occupation numbers $n$. }
		\label{tab:basis}
\end{table}

\subsection{Ladder dynamical vertex approximation}
Within the D\textGamma A approach, two- and four-point Green's functions are obtained directly \cite{rohringer_diagrammatic_2018}. However, the above correlators require higher order Green's functions at equal times. As explained in \cite{Roosz_2024_2sRDM} and applied in \cite{Bippus_2sRDM_entanglement_2026}, these can be obtained by leveraging the Heisenberg equation of motion \begin{equation}
    \frac{\partial \hat c_{i\sigma}}{\partial \tau} = [H,\hat c_{i\sigma}] = t\sum_{j(i)}\hat c_{j \sigma} + \mu \hat c_{i \sigma} - U \hat c^{\dagger}_{i\overline{\sigma}}\hat c^{\phantom \dagger}_{i\overline{\sigma}}\hat c_{i\sigma},
\end{equation}
which crucially includes a term dependent on three fermionic operators. This term allows for the reduction of the above equal-time correlators to four-point quantities by the introduction of imaginary-time derivatives. These, in turn, can be Fourier transformed into sums over Matsubara frequencies, which all can be computed from the normal D\textGamma A results. In compliance with the full parquet D\textGamma A approach \cite{Bippus_2sRDM_entanglement_2026,Toschi2007,Valli2015,rohringer_diagrammatic_2018,krien_tiling_2021,krien_plain_2022}, ladder quantities must be rewritten into full momentum dependencies via the crossing relation \cite{rohringer_diagrammatic_2018}. 
Due to the limited numerical precision, slightly negative eigenvalues of the reduced density matrix occur. These negative eigenvalues are corrected by a unity trace-preserving optimization to the nearest physical density matrix \cite{Bippus_2sRDM_entanglement_2026,smolin_efficient_2012}.

\section{Full negativity}\label{Full_N}
\begin{figure}
    \centering
    \includegraphics[width=0.5\linewidth]{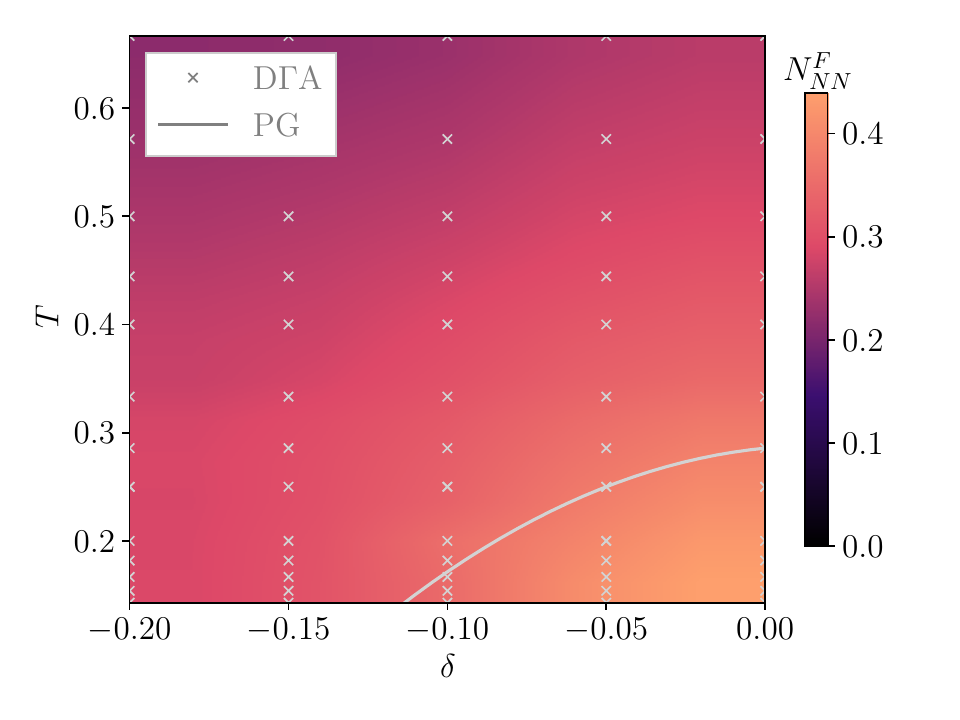}
    \caption{The nearest neighbor full fermionic negativity $N^F$ as measured in D$\Gamma$A as a function of temperature $T$ and electron doping $\delta$. The onset of the pseudogap regime is marked by the white dashed line.}
    \label{fig:NF}
\end{figure}

In bosonic systems the partial transpose acting on any operator $O$ has a clear definition 
\begin{equation}\label{partial_transpose}
    \langle \psi_i, \psi_j | O^{T_j} | \bar{\psi}_i, \bar{\psi}_j \rangle \equiv  \langle \psi_i,  \bar{\psi}_j |O | \bar{\psi}_i, \psi_j \rangle,
\end{equation}
where $\psi_{i/j},\, \bar{\psi}_{i/j}$ are any two states from Table~\ref{tab:basis} \cite{Peres_1996,Horodecki_1996}. However, this neglects the fermionic signs introduced by the swapping of states from different parity sectors. When applying the SSR rules, states for which this would be the case, are naturally set to zero. Exactly those elements and signs are crucial however for identifying the full entanglement in a system (note that this entanglement is then not accessible by local measurements any longer and while it carries physical information \cite{Bippus_2sRDM_entanglement_2026} it can not contribute to any meaningful quantum information theory). To reinstate the fermionic signs, one introduces the partial time-reversal operator \cite{Shapourian_2017_first_NF,Shapourian_2019_second_NF, Bippus_2sRDM_entanglement_2026} 
\begin{equation}
    \langle \psi_i, \psi_j | \hat O^{{\rm TR}_j} | \bar{\psi}_i, \bar{\psi}_j \rangle \equiv  \langle \psi_i,  \bar{\psi}_j |\hat O | \bar{\psi}_i, \psi_j \rangle (-1)^{\phi}
\end{equation}
with the phase factor 
\begin{equation}
\begin{aligned}
\phi = & \frac{{\cal N}_i\left({\cal N}_i+2\right)}{2}+\frac{\bar{{\cal N}}_i\left(\bar{{\cal N}}_i+2\right)}{2}+\bar{{\cal N}}_j {\cal N}_j \\
& +{\cal N}_i {\cal N}_j+\bar{{\cal N}}_i \bar{{\cal N}}_j+\left(\bar{{\cal N}}_i+\bar{{\cal N}}_j\right)\left({\cal N}_i+{\cal N}_j\right)
\end{aligned}
\end{equation}
defined through the number of electrons ${\cal N}_i = \langle \hat n_{i,\uparrow}\rangle + \langle \hat n_{i,\downarrow}\rangle$ of each matrix element.
This then gives rise to the fermionic negativity $N^F$ defined in full analogy to Eq.~7:
\begin{equation}\label{eq_negativity_SM}
    N^F = \log_2\left( \| \rho_{ij}^{{\rm TR}_j} \|_{\textrm{tr}} \right).
\end{equation}
As mentioned in the main text, the fermionic negativity detects trivial entanglement even outside of the pseudogap regime. For $U=6.5t$ and various dopings we present the theoretical result for nearest neighbor lattice sites in Fig.~\ref{fig:NF}.
This entanglement also exists at low values of the interaction $U$ and even at larger distances \cite{Bippus_2sRDM_entanglement_2026}. Therefore, we want to stress here, that the detected entanglement is not characteristic specifically for the pseudogap regime and note again that this entanglement is induced by electrons hopping through the lattice which is not measurable by local operations.

\section{Error estimation and non-existence of SSR entanglement beyond nearest neighbors}
\begin{figure}
    \centering
    \includegraphics[width=1\linewidth]{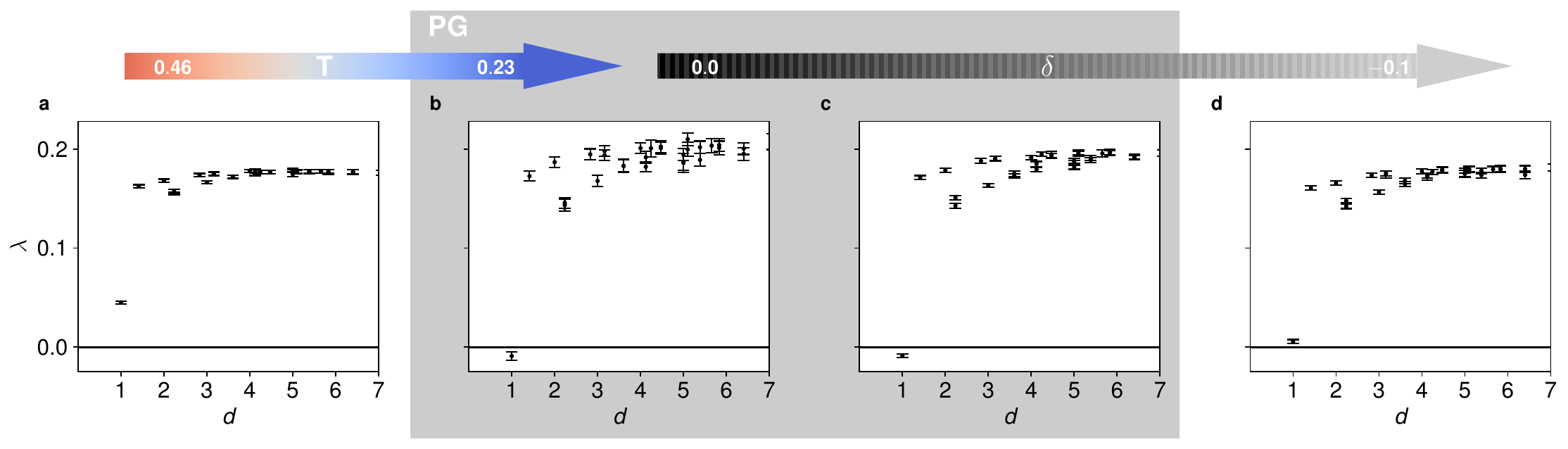}
    \caption{Relevant eigenvalue $\lambda$ of the partial transposed density matrix for the SSR negativity for different temperatures $T$ and electron doping $\delta$ over distance $d$ from experiment, including error bars. A SSR negativity exists for values smaller zero. The grey box marks data points within the pseudogap regime.}
    \label{fig:eig_cond}
\end{figure}

A SSR negativity and hence spin-singlet entanglement is only observed if the partial transposed SSR filtered density matrix yields a negative eigenvalue. Hence, the relevant eigenvalue $\lambda$ from Eq.~\eqref{cond_rho} needs to be negative. In Fig.~\ref{fig:eig_cond}, we show this eigenvalue for different values of doping and temperatures over distance $d$. Error bars are obtained by Gaussian error propagation of the bootstrapping errors obtained for the relevant density matrix elements (see \cite{Chalopin_PG_2026} for more details). Only within the pseudogap regime, the nearest neighbor eigenvalue becomes negative within the experimental error bars and gradually shifts towards positive values in the crossover to the metallic regime. For any distance larger $d=1$, the relevant eigenvalue stays significantly above zero. Since the Peres-Horodecki criterion is exact for small systems, such as the SSR sector we investigate here, a zero SSR negativity proves that there is no spin-singlet entanglement in the system \cite{Horodecki_1996}. Hence, our experimental observation shows that spin-singlet entanglement is restricted to the pseudogap regime and nearest neighbour lattice sites only. This observation is in agreement with our theoretical D$\Gamma$A result. Again, we refer the reader to Supplemental Information~\ref{Full_N} for a discussion of operational inaccessible entanglement going beyond spin-singlets.

\bibliography{references_fixed.bib}